\documentclass[11pt,a4paper,final]{iopart}
\usepackage[utf8]{inputenc}
\usepackage{graphicx}
\usepackage{bm}
\usepackage{rotating}
\usepackage{multirow}

\usepackage{hyperref}
\usepackage[normalem]{ulem}
\usepackage{comment}
\usepackage{iopams}
\usepackage{titlesec}
\expandafter\let\csname equation*\endcsname\relax
\expandafter\let\csname endequation*\endcsname\relax
\usepackage{amsmath}
\usepackage{amssymb}
\def\orcid#1{\kern .08em\href{https://orcid.org/#1}{\includegraphics[keepaspectratio,width=0.7em]{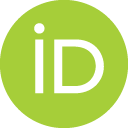}}}
\usepackage[belowskip=-5pt]{caption}
\usepackage{pifont}
\usepackage{dcolumn}
\usepackage{bm}
\usepackage{float}
\usepackage[switch*]{lineno}
\usepackage{xcolor}
\usepackage{adjustbox}
\usepackage{ctable}
\usepackage[bottom]{footmisc}

\usepackage{appendix}

\begin{document}

\title{Connecting ground-state properties of ${}^6$Li to each other and to scattering data}

\author{C. Hebborn$\,\,\!\!$\orcid{0000-0002-0084-2561}$^{1,2,3,*}$, C. R. Brune$\,\,\!\!$\orcid{0000-0003-3696-3311}$^{4,\dagger}$, D. R.  Phillips$\,\,\!\!$\orcid{0000-0003-1596-9087}$^{4,5,\ddagger}$}

\vspace{0.3cm}
\address{$^1$Université Paris-Saclay, CNRS/IN2P3, IJCLab, 91405 Orsay, France.}
\address{$^2$Facility for Rare Isotope Beams, Michigan State University, East Lansing, Michigan 48824, USA.}
\address{$^3$Department of Physics and Astronomy, Michigan State University, East Lansing, Michigan 48824, USA.}
\address{$^4$Department of Physics and Astronomy and Institute of Nuclear and Particle Physics, Ohio University, Athens, Ohio 45701, USA.}
\address{$^5$Department of Physics, Chalmers University of Technology, SE-41296 G\"oteborg, Sweden. }

\vspace{0.3cm}

\address{E-mails: $^{*}$hebborn@ijclab.in2p3.fr, $^{\dagger}$brune@ohio.edu, $\ddagger$phillid1@ohio.edu}

\begin{abstract}
  We examine the relationship between the Asymptotic Normalization Coefficient (ANC) of $^6$Li and other low-energy observables in the $\alpha$-deuteron system. Our analysis  uses a set of calculations carried out within the {\it ab initio} No Core Shell Model with Continuum (NCSMC) using a variety of inter-nucleon 
 interactions and basis sizes, and yielding ${}^6$Li deuteron separation energies between 1.3 and 1.8 MeV~\cite{PhysRevLett.129.042503}. These NCSMC calculations show that the square of the ANC is strongly correlated with the separation energy over this range. In this work, we investigate the origin of  this correlation using  the phenomenological $R$-matrix, a single-channel potential and a perturbative approach. We show that this correlation occurs because the depth of the $\alpha$-deuteron central potential changes by only a small relative amount as the separation energy varies. We then investigate if the ANC can be accurately extracted from  $\alpha$-deuteron phase shifts in an ideal case in which low-energy data are available and there are no experimental errors. We find that both $R$-matrix and Coulomb-modified effective-range theory (CM-ERE) yield  extracted ANCs  close to, although not exactly equal to, the true value, 
 provided the extrapolation is constrained by the known position of the bound-state pole  and at least three terms are included in the fit function. The $R$-matrix approach converges faster than the CM-ERE as the number of parameters increases and is also more robust against the inclusion of low-energy and high-energy phase shift data. Finally, our study also shows that a naive quantification of uncertainties by comparing different truncations used in both theories is not accurate, and suggests the accuracy of  ANCs extracted from phase shift data needs further investigation.
     \end{abstract}
\maketitle

\section{Introduction}
Our understanding of the origins of the elements  in the Universe relies on nuclear reaction networks, that take as inputs evaluations of low-energy nuclear reaction rates. Because they provide a path to the formation of heavier elements,  radiative capture reactions, in which two nuclei $A$ and $B$ fuse together to produce a heavier nucleus $C$ and a gamma-ray, are of particular interest for various  nucleosynthesis processes.  These low-energy capture reactions depend on the overlap function $u(r)$ of the ground state of $C$ with the ground states of $A$ and $B$ and the $A$-$B$ scattering wavefunctions. 
At large $A$-$B$ distance $r\to \infty$, these overlap functions $u(r)$  depend on two physical quantities: $E_b$ and the Asymptotic Normalization Coefficient (ANC) of the state.
 ANCs are  model-independent observables, that can be obtained formally from the residue of the $A+B$ $S$-matrix at the bound state pole $C$~\cite{Locher1978,Plattner1981,PhysRevC.37.2859,PhysRevC.48.2390,PhysRevC.59.3418,Yarmukhamedov:2011kd,PhysRevC.95.044618,Muk24}. Because many evaluations of astrophysical rates rely on these ANCs, it is important to understand how they correlate with other observables that  can be measured, i.e., scattering cross sections and the nuclear binding energy.

This paper focuses on the $s$-wave $\alpha$-$d$ ANC of $^6$Li. This ANC determines a sizable fraction of the cross section for the radiative capture $\alpha + d \rightarrow ^6{\rm Li} + \gamma$, and so is strongly implicated in the rate of ${}^6{\rm Li}$ production during the Big Bang. We analyze the recent state-of-the-art {\it ab initio} predictions carried of Hebborn {\it et al.}  which used a variety of nuclear interactions derived from chiral effective field theory ($\chi$EFT)~\cite{PhysRevLett.129.042503}. Hebborn {\it et al.} found that the capture rate varied very little among  different interactions that reproduced the separation energy, and traced this result to the stability of the $^6$Li $s$-wave  ANC $C_0$ across those different \textit{ab initio} calculations. Thanks to the stability of the many-body predictions with respect to the choice of interactions and model spaces, they predicted a precise value for the ANC$^2$ of $^6$Li: $C^2_0=6.86 \pm 0.21 $~fm$^{-1}$.

As discussed in Refs.~\cite{PhysRevLett.129.042503,Hebborn:2023bwu}, this recent \textit{ab initio}  ANC differs  from the accepted ANC$^2$ value $C_0^2=5.3\pm 0.5 $~fm$^{-1}$ obtained by analytic continuation of $s$-wave phase shifts, from an energy-dependent phase-shift analysis~\cite{PhysRevC.43.822} that was cross-checked against both a single-particle reaction model constrained with experimental phase shifts and a three-body $\alpha$+$n$+$p$ model for the structure of ${}^6$Li that neglects the $\alpha$-$p$ Coulomb repulsion~\cite{PhysRevC.48.2390}. This ANC discrepancy does not just have consequences for the evaluation of the $\alpha + d \rightarrow ^6{\rm Li} + \gamma$ capture rate. The ${}^6$Li ANC has been used to analyze $({}^6{\rm Li},d)$ transfer data to put constraints on $\alpha$-induced reactions of astrophysical interest, such as  ${}^{13}$C$(\alpha,n) {}^{16}$O~\cite{PhysRevC.91.048801}  and ${}^{12}$C$(\alpha,\gamma) {}^{16}$O~\cite{PhysRevLett.83.4025,PhysRevLett.114.071101}. Thus a revision of the ${}^6$Li ANC  means that these  reaction rates should also be  reevaluated~\cite{Hebborn:2023bwu}.
In this work, we investigate this discrepancy in ANCs of $^6$Li from two different perspectives. First, we analyze the strong correlation between the ${}^6$Li deuteron separation energy and the square of the ${}^6$Li ANC. Second, we study how accurately this quantity can be determined when phase-shift data is used to constrain an extrapolation of the scattering amplitude to the bound-state pole. 

The correlation between binding energy and ANC has already been observed in other few-nucleon systems, such as ${}^3$He and the triton~\cite{PhysRevC.37.2859}. 
For three-nucleon systems, these correlations  are expected at leading order in pionless effective field theory (EFT)~\cite{Weinberg:1991um,vanKolck:1998bw,Chen:1999tn}. Indeed, the presence of a single three-nucleon contact interaction in the leading-order Lagrangian of that theory~\cite{Bedaque:1998kg,Bedaque:1999ve} implies that all three-nucleon (and some four-nucleon) observables are correlated---at least up to corrections that are sub-leading in this theory~\cite{Bedaque:2002yg,Bazak:2018qnu}. Thus, in pionless EFT, the ANC-separation energy correlation observed in Ref.~\cite{PhysRevC.37.2859} has the same explanation as the Phillips and Tjon lines~\cite{Bedaque:1999ve,Phillips:1968zze,Tjon:1975sme,Platter:2004zs}. ${}^6$Li was computed in pionless EFT in Ref.~\cite{Stetcu:2006ey} and
a  correlation between $E_b$ with the $\alpha$-deuteron scattering length~\cite{Lei:2018toi} has  been observed in a recent study employing a three-body model  (but without Coulomb)~\cite{Lei:2018toi}.
In this work, we discuss  a slightly different although similar correlation, between the ANC$^2$ and the separation energy of $^6$Li in many-body calculations using not pionless EFT, but nuclear forces derived from chiral EFT.
A broader understanding of what causes such correlations could reveal for which systems they occur and might help to improve evaluations of astrophysical reaction rates. 

Since  the deuteron separation energy in ${}^6$Li is small compared to the typical binding energy per nucleon in nuclei, Halo EFT~\cite{Hammer:2017tjm} is a natural framework 
for describing the ${}^6$Li  system. Halo EFT uses this separation of energy scales to systematize the interactions between the $\alpha$ and the deuteron clusters. Observables can then be derived up to a given order in the small momenta associated with the separation energy. 
The form of the scattering amplitude that emerges in Halo EFT is that of  the Coulomb-modified Effective Range Expansion (CM-ERE)~\cite{PhysRev.77.647,Sparenberg}, however with a different organization of the CM-ERE in terms of the dimensionful scales.
In particular, at leading order (LO) in Halo EFT for s-wave states only one term appears in the CM-ERE and hence s-wave states exhibit a parameter-free relationship between the two-body separation energy and many bound-state properties, such as the radius, electromagnetic matrix elements and the ANC~\cite{Hammer:2017tjm}.
For a weakly bound state without Coulomb interactions the ANC$^2$ at LO can simply be obtained from the binding momentum; if the bound state involves two charged particles this dependence is modified by an extra factor depending on the Sommerfeld parameter of the bound state~\cite{Kong:1998sx,Higa:2008dn,Ryberg:2015lea}. 
One might think that these simple relations at LO can explain  the correlation between binding energy and ANC$^2$, and that one can interpret this correlation as a manifestation of  ``universality" in $^6$Li, but we will see below that this is not the case. 

Because it was recently argued that ${}^6$Li could be treated as  a single-channel problem within Halo EFT of ${}^6$Li~\cite{Nguyen:2025nvu}, we also investigate the ANC-binding energy correlation using a single-particle model and the phenomenological $R$-matrix framework. $R$-matrix is particularly suited to studying shallow bound states, since it explicitly separates the long-distance piece of the bound-state wave function from the interior portion that is treated phenomenologically.  In the first half of this work, we  compare $R$-matrix, single-particle model, and perturbative arguments for the correlation between the binding energy and the ANC of of ${}^6$Li. %
In the second half, we use both $R$-matrix  and the CM-ERE theories to explore the connection between phase shifts and the ANC. 
We focus on the $1^+$ channel in which the bound-state occurs and convert the NCSMC phase shifts into a parametrization of the inverse scattering amplitude. We then take the inverse scattering amplitude between 0.01 MeV and 3 MeV as data and fit CM-ERE and $R$-matrix amplitudes to those data. We  view this as a first step towards investigating the robustness of ANC extractions from experimental data.

Our paper is structured as follows. In Sec.~\ref{sec:NCSMCdetails} we give details of the NCSMC calculations of ${}^6$Li that form the basis of our study. Sec.~\ref{sec:ANCBEconnection} exposes the correlation between the separation energy and the ANC squared from three different perspectives. In Subsec.~\ref{subsec:Rmatrix} we explain the existence of such a correlation from an $R$-matrix perspective, then in Subsec.~\ref{subsec:WoodsSaxon} we show it also emerges in a single-particle potential model. These two perspectives are linked by their common assumption that the change in separation energy is small compared to the overall depth of the two-body potential, and  Subsec.~\ref{subsec:perturbative} shows that affine dependence of ANC$^2$ on binding energy is to be expected in this regime.
In Sec.~\ref{sec:ANCPSconnection}, we turn out attention the connection between phase shifts and the ANC. After a review of the CM-ERE in Subsec.~\ref{subsec:CMEREreview} we present results from CM-ERE fits to the positive-energy NCSMC scattering amplitude in Subsec.~\ref{subsec:CMEREresults}. Subsec.~\ref{subsec:Rmatrixresults} then carries out the same fitting exercise using the $R$-matrix form of the amplitude. We summarize and offer perspectives on the extension of our discussion of ${}^6$Li to other systems in Sec.~\ref{sec:summary}. 

\section{Description ab initio calculations considered}
\label{sec:NCSMCdetails}

In this work, we analyze calculations of bound $^6$Li and scattering $\alpha$-$d$ states  that were carried out in Ref.~\cite{PhysRevLett.129.042503} using the  \textit{ab initio}  no-core shell model with continuum (NCSMC) method~\cite{PhysRevC.87.034326,PhysRevLett.110.022505}. This \textit{ab initio} approach has the advantage that it treats accurately and consistently static and dynamic properties of light nuclei (see Refs.~\cite{Navratil_2016, Navratil2020, QUAGLIONI2025123095} for recent reviews). This is done through an expansion onto a basis composed of  $^6$Li no-core shell model (NCSM) states as well as  $\alpha$-$d$ cluster states, with the latter built from  $\alpha$ and $d$ NCSM states.   The NCSMC calculations use interactions derived from chiral effective field theory ($\chi$EFT), but consider several such interactions. Some employ only NN interactions and some use different 3N $\chi$EFT forces~\cite{PhysRevLett.103.102502,PhysRevLett.122.029901,PhysRevC.101.014318}.  The Hamiltonians are softened to a variety of momentum resolution scales using similarity renormalization group  (SRG) transformations~\cite{PhysRevLett.103.082501}. The calculations also consider different truncations of the basis used for diagonalizing the $\chi$EFT Hamiltonian: the results are obtained  including a different numbers of particle excitation quanta ($N_{max}$) above the lowest-energy configuration  and with different numbers of deuteron states, i.e., including  the deuteron ground state and pseudostates, thereby effectively discretizing the $n$-$p$ continuum. Finally, predictions with  and without a phenomenological adjustment  of the energy of the $\alpha$-$d$ separation threshold are also considered. Twelve different calculations, that produced ${}^6$Li states bound by 1.39--1.86 MeV with respect to the $\alpha-d$ threshold, will be studied in this work. Table~\ref{TableNCSMC} provides a summary of the ingredients in each one.  

    \begin{table}
	\centering	
	\small\begin{tabular}{l|ccccc||c}
		{number}&$\lambda$ [fm$^{-1}$]&	with 3N & with ps & Pheno& $N_{max}$& {$E_b$ [MeV]}\\ \hline\hline
		0&	2.00&$\checkmark$&	$\times$&$\times$&10&-1.75\\
		1&2.00&$\checkmark$	& $\times$		&$\checkmark$	&10&-1.48\\
		2&2.00&		$\times$		& $\checkmark$&		$\times$&8	&-1.86\\
		3&2.00&	$\times$&	$\checkmark$&	$\times$&10&-1.85\\
		4&2.05&	$\times$&$\checkmark$&$\times$&8&-1.75\\
		5&2.20&	$\times$& $\checkmark$&$\times$&8&-1.46\\
		6& 2.00&$\times$&$\checkmark$&$\checkmark$&8&-1.48\\
		7& 2.00&$\checkmark$&$\checkmark$&$\times$ &8&-1.84\\
        8&2.00&$\checkmark$&$\checkmark$&$\times$ &10&-1.78\\
		9&2.00&		$\checkmark$&$\checkmark$&$\checkmark$&10&-1.48\\
		10&	2.00&	$\checkmark$&$\checkmark$&$\times$&8&-1.39\\
		11&	2.00&	$\checkmark$&$\checkmark$&$\checkmark$&8&-1.47\\
	\end{tabular}
 \caption{Summary of the \textit{ab initio} calculations analyzed in this work. The first column gives the number by which we refer to the calculations in subsequent text and figures, the second is the momentum resolution scale of the SRG evolution ($\lambda$), while the third and fourth indicate if the 3N chiral  (and SRG-induced) force was included and if deuteron pseudostates were included. The fifth column shows whether a phenomenological adjustment of the threshold energy was done {\it a posteriori}, and the sixth gives the number of quanta  ($N_{max}$) of excitation above the lowest-state energy. The last column is then the result of that NCSMC calculation for the deuteron separation energy of the ${}^6$Li ground state. In all cases, the NN force considered is from Ref.~\cite{PhysRevC.68.041001}. Two 3N forces were considered: from Ref.~\cite{PhysRevLett.103.102502,PhysRevLett.122.029901} for predictions 0, 1, 7, 8 and 9  and  from Ref.~\cite{PhysRevC.101.014318} for predictions 10 and 11.} \label{TableNCSMC}
\end{table}

Our analysis focuses on the asymptotic properties of the overlap of the $^6$Li ground state and scattering states formed out of a direct product of the ground states of the $\alpha$-particle and the deuteron. The dominant  overlap is  of $s$-wave character~\cite{PhysRevLett.129.042503}. Asymptotically, the $s$-wave radial overlap function $u_0(r)$ is proportional to the Whittaker $W$ function
\begin{equation}
    u_0(r)  \xrightarrow[r \to \infty]{} C_{0} W(-\eta_b,\frac{1}{2},2 \kappa r), \label{eqBound}
\end{equation}
where $r$ is the $\alpha$-$d$ relative distance, $\kappa=\sqrt{2\mu |E_b|}/\hbar$ is the binding momentum, and
${C}_0$ the s-wave ANC. $\eta_b=\kappa/k_C$  is the Sommerfeld parameter obtained from the binding momentum $\kappa$ and $k_C=Z_1 Z_2 \alpha_{\rm em} \mu$, with $\mu$ the reduced mass of the two-body system and $Z_1=2$ and $Z_2=1$ the charges of, respectively, the $\alpha$ and $d$ nuclei in units of $|e|$. Because the NCSMC explicitly includes $\alpha$-$d$ cluster states in the basis, the ANC is  directly obtained from matching the calculated wavefunction with the asymptotic behavior~\eqref{eqBound}~\cite{Navratil_2016,Navratil2020,QUAGLIONI2025123095}. In Section~\ref{sec:ANCBEconnection}, we investigate the correlation between $C_0^2$  and deuteron separation energy $E_b$ obtained in the various \textit{ab initio }calculations.

In the scattering regime, we are interested in the phase shifts, which dictate the asymptotic behavior of the overlap function for positive energy. In a single-channel $s$-wave framework, the asymptotic behavior of these scattering wave functions at an energy $E$ is
\begin{equation}
    u_0(r)  \xrightarrow[r \to \infty]{} H_0^- (\eta, kr)+e^{2i\delta_0} H^+_0(\eta, kr),
\end{equation}
where $H_0^{\pm}$ are the outgoing (+) and incoming (-) Coulomb Hankel functions, defined for relative  angular momentum $\ell=0$, $\delta_0$ is the corresponding phase shift, while $k$  and $\eta$ are  the wavenumber and Sommerfeld parameter corresponding to the energy $E$~\footnote{To connect the momenta and Sommerfeld parameters for the scattering and bound states the momentum $k$ is analytically continued to $k=i \kappa$, and $\eta$ is correspondingly continued to $\eta=-i \eta_b$, $\eta_b \equiv k_C/\kappa$.}. In Section~\ref{sec:ANCPSconnection}, we investigate the connection of the $s$-wave phase shift with $C_0$.

\section{The ANC-binding energy connection}
\label{sec:ANCBEconnection}

We first investigate the connection between the binding energy $E_b$ and the $s$-wave ANC, $C_0$. As seen in Eq.~\eqref{eqBound}, only these two quantities are relevant for the asymptotic part of the bound-state wavefunction. The first  dictates how steeply the wavefunction decreases, while the second gives the overall  normalization of this wavefunction at large distance. 

In Fig.~\ref{fig:scaling_rmatrix}, we plot the $s$-wave ANC$^2$s obtained with the various NCSMC calculations, as a function of the binding energy (black points). Although  these predictions result from various complex \textit{ab initio} calculations (see Table~\ref{TableNCSMC}), this figure suggests the existence of a strong correlation. The correlation ensures that different Hamiltonians and model-space truncations in Table~\ref{TableNCSMC} which reproduce the empirical ${}^6$Li binding energy have only a small spread in their $^6$Li s-wave ANCs. This provides an explanation of the small uncertainty in the NCSMC calculation of the $\alpha(d,\gamma)^6$Li reaction rate~\cite{PhysRevLett.129.042503}.

In this section,  we analyze this correlation using different frameworks: the phenomenological $R$-matrix framework, a single-particle potential model, and  first-order perturbation theory. In order to simplify the discussion below, we define here some useful quantities. The $s$-wave $\alpha$-$d$ NCSMC overlap function is taken to be $u_0(r)$, which can be chosen to be real assuming time-reversal invariance. We then define	
\begin{align}
I_{\rm in}(\tilde{R}) &= \int_0^{\tilde{R}}| u_0(r)|^2 \, dr , \label{eq:interiornorm}\\
I_{\rm out}(\tilde{R}) &= \int_{\tilde{R}}^\infty |u_0(r)|^2 \, dr \mbox{,~and} \\
I_{\rm as}(\tilde{R}) &= \int_{\tilde{R}}^\infty |W(-\eta_b,\frac{1}{2},2 \kappa r)|^2 \, dr .
\end{align}
Note that all three integrals depend on the binding momentum $\kappa$. Finally, we also define $N_0=I_{\rm in}+I_{\rm out}$, which is independent of $\tilde{R}$ and is the spectroscopic factor of the $s$-wave $\alpha$-$d$ configuration in the bound state of ${}^6{\rm Li}$.

\subsection{Phenomenological $R$-matrix theory}
\label{subsec:Rmatrix}

The phenomenological $R$-matrix theory is a powerful method, commonly used to analyze low-energy cross sections, relevant for astrophysics~\cite{RevModPhys.30.257,Descouvemont_2010,RevModPhys.89.035007,msrx-3fjr}.  It  relies on a division of coordinate space into an internal and external region, with the transition occurring at a certain channel radius $a$. The spectrum of the Hamiltonian in the region $r < a$ defines a set of energies $E_n$ and a corresponding 
set of eigenfunctions $\phi_n$. The $R$-matrix can then be expressed in terms of these quantities. The $R$-matrix thus replaces the solution to the Hamiltonian problem for $r<a$, and properties of the system can be calculated by matching the properties of the $R$-matrix at $r=a$---its levels $E_n$ and the levels' partial widths $\gamma_n$---with the asymptotic solution. 

In the phenomenological $R$-matrix approach, the levels $E_n$ and partial widths $\gamma_n$ are fitted to two-body scattering data, one can then infer properties of  states or make predictions for cross sections at energies where no experimental data are available.  In the $1^+$ partial wave at low energies, $^6$Li is known to be dominated by one channel, $s$-wave $\alpha$-$d$, so the $R$-matrix formalism provides a natural framework to study the ANC-binding energy relation, as well as the phase shift-ANC connection. Some care is required though, as  typically, one takes a channel radius around 3--5~fm, and the asymptotic form of the wavefunctions is not completely reached there.

For a single-channel $s$-wave problem, the relationship between $\gamma$ and the ANC$^2$ is~\cite{Plattner1981,PhysRev.84.1061,PhysRevC.66.044611,PhysRevC.102.034328}.
\begin{equation} \label{eq:anc}
C_0^2 = \frac{2\mu a}{\hbar^2|W(-\eta_b,l+\frac{1}{2},2 \kappa a)|^2}\frac{\gamma^2}{1+\gamma^2 \left. \frac{dS_0}{dE} \right|_{E=-E_B}},
\end{equation}
where 
 $S_0$ is the $s$-wave shift function, which for negative energies is
\begin{equation}\label{eq:shift}
S_0=\left(\frac{r}{W(-\eta_b,\frac{1}{2},2 \kappa r)}\frac{dW(-\eta_b,\frac{1}{2},2 \kappa r)}{dr}\right)_{r=a}.
\end{equation}
$dS_0/dE$ can be expressed as~\cite[Eq.~(29), p.~351]{RevModPhys.30.257}
\begin{equation} \label{eq:dsde}
\frac{dS_0}{dE} = \frac{2\mu a}{\hbar^2 |W(-\eta_b,\frac{1}{2},2\kappa a)|^2} I_{\rm as}(a).
\end{equation}
A correlation between the ANC$^2$ and the binding energy can then be straightforwardly derived in the phenomenological $R$-matrix approach. Substituting Eq.~(\ref{eq:dsde}) into Eq.~(\ref{eq:anc}) we have:
\begin{equation}
    \label{eq:Rmatscaling}
    C_0^2=\left[\frac{1}{\gamma^2} \frac{\hbar^2 |W(-\eta_b,\frac{1}{2},2\kappa a)|^2}{2 \mu a} + I_{\rm as}(a)\right]^{-1}.
\end{equation}
If changes in the interaction are small compared to the expectation values of the potential and kinetic energies, then the change in the bound state wave function is
likewise small. Then, inside the channel radius, i.e., for $r<a$, the wave function should hardly change.
 In the phenomenological $R$-matrix approach this means the reduced width, $\gamma$, should remain constant. 

Since we have direct access to the overlap function in the NCSMC calculation, we can check the extent to which $\gamma^2$ is independent of $E_b$. 
Using~\cite[Eqs.~(56) and~(64)]{PhysRevC.102.034328}
\begin{equation}
    \gamma^2 =\frac{\hbar^2}{2\mu a} \frac{|u_0(a)|^2}{1-I_{\rm out}(a)} . \label{eq:reducedwithd}
\end{equation}
for NCSMC calculation numbers 2, 3, 6, 8, 9, and 11 (see Table~\ref{TableNCSMC}) produces $\gamma^2$ values that are within 13\% of each other.

\begin{figure}
    \centering
    \includegraphics[width=0.7\linewidth]{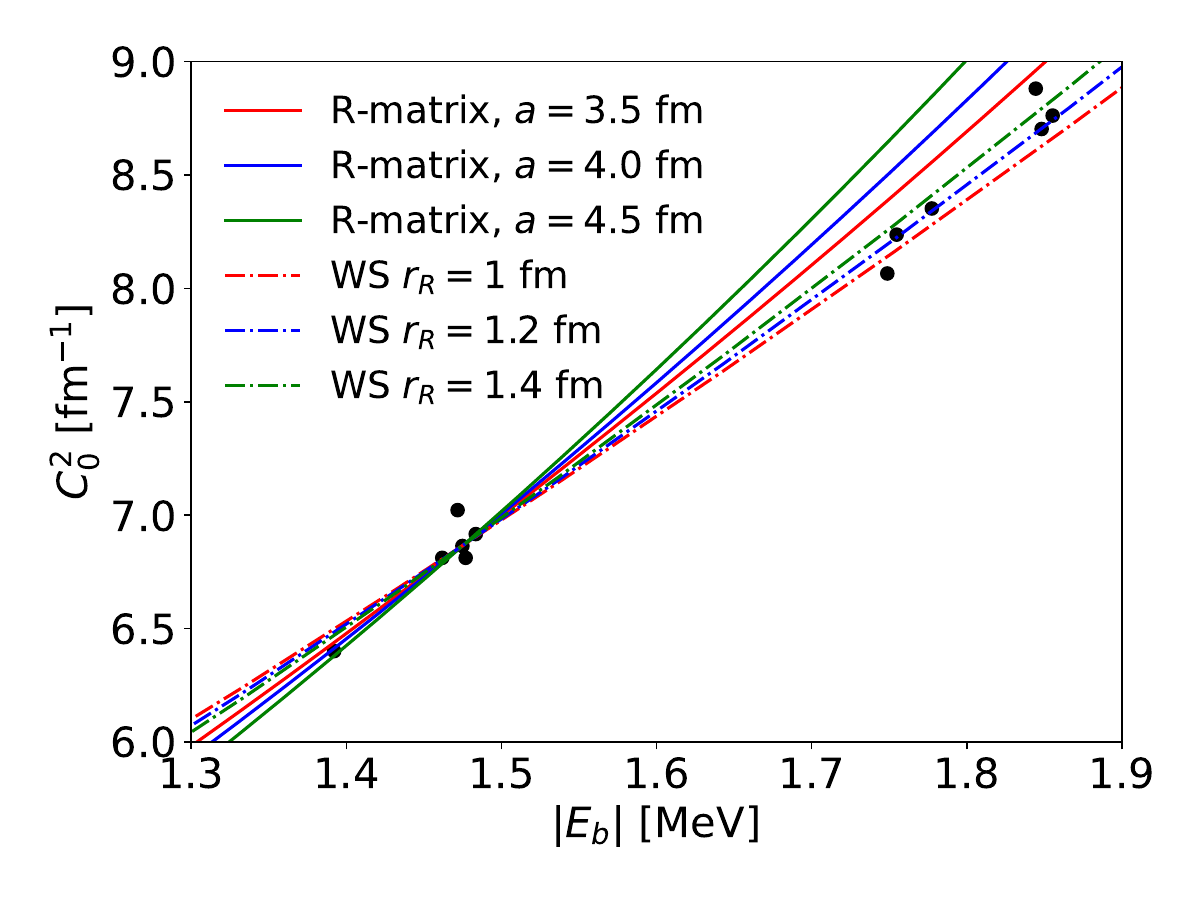}
    \caption{ANC$^2$s versus the $\alpha$-$d$ separation energy. A strong correlation is indicated by the NCSMC calculations (black points). The correlation predicted by the $R$-matrix approach~\eqref{eq:anc} for three different values of the channel radius $a$ is shown as the solid curves. The dot-dashed curve shows the correlation predicted using a Woods-Saxon (WS) potential.}
    \label{fig:scaling_rmatrix}
\end{figure}

If we neglect this small variation in $\gamma^2$ with $\kappa$ then Eq.~(\ref{eq:Rmatscaling}) predicts that the binding-energy dependence of the ANC$^2$ comes solely from the energy dependence of the Whittaker function (including in $I_{\rm as}(a)$). The resulting prediction is shown for $a=3.5$~fm, $a=4$~fm and $a=4.5$~fm (solid lines) in Fig.~\ref{fig:scaling_rmatrix}. Here the value of $\gamma$ was fixed to reproduce NCSMC prediction 9 for the ANC and binding energy (see Table~\ref{TableNCSMC}). The $R$-matrix calculation describes the correlation between the binding energy and the ANC$^2$s reasonably well. 

Note that Eq.~\eqref{eq:reducedwithd} is equivalent to Eqs.~\eqref{eq:anc}-\eqref{eq:dsde}, if $u_0(r)=C_0 W(-\eta_b,\frac{1}{2},2\kappa r)$ for $r\ge a$, i.e., if the overlap function is fully asymptotic beyond the channel radius. 
This indicates that reduced-width amplitudes deduced using Eq.~\eqref{eq:anc} 
will not be directly obtainable from {\it ab initio} calculations if $a$ is not in the asymptotic region.
However, one does not want to make the channel radius much larger than the ones employed in the previous paragraph. Doing so will cause $\gamma$ to change with $\kappa$ because of the changing exponential decay outside the nucleus.

In fact, though, the binding-energy dependence of the NCSMC overlap functions in the interior region is very similar to the dependence in the asymptotic region. Ratios of the overlap functions at different binding energies
\begin{equation}
\frac{u^{(1)}_{0}(r)}{u^{(2)}_{0}(r)} = \frac{C^{(1)}_{0}W(-\eta^{(1)}_{b},\frac{1}{2},2\kappa^{(1)} r)}{C^{(2)}_{0}W(-\eta_b^{(2)},\frac{1}{2},2\kappa^{(2)} r)},
\label{eq:clusterfnscaling}
\end{equation}
are independent of $r$ for $r > 2.5$~fm. Here the superscripts 1 and 2 refer to two different NCSMC calculations that produce two different binding energies. This explains the success of the $R$-matrix scaling relation, in spite of the fact that the overlap function is not asymptotic for some of the channel radii used.

\subsection{Single-particle potential models}

\label{subsec:WoodsSaxon}

To further understand why these NCSMC reduced widths stay rather constant across various predictions, considering different Hamiltonian and model spaces, we study the simple case of  a finite square well of radius $a_{sw}$ and depth $V_0$. In this toy model, one can easily determine analytically the bound state wavefunction and impose that it reproduces various binding energies by adjusting the depth of the well. From these solutions,  the corresponding reduced widths can be obtained using~\eqref{eq:reducedwithd}. 
We chose $a_{sw}=2$~fm and tuned the well so that its first excited state occurred at energies from $-1.9$~MeV to $-1.3$~MeV. We found the depth of the potential needs to be changed very little in order to achieve this: it moves only from $\approx$ 93.5~MeV to $\approx$ 95 MeV. Meanwhile $\gamma^2$ undergoes an even weaker binding-energy dependence, changing only from 7.35 to 7.30. However, $C_0^2$ changes markedly, it spans a range from $2.25~{\rm fm}^{-1}$ to $3~{\rm fm}^{-1}$. 
This additional analysis traces back the  weak binding-energy dependence of the partial width to the weak binding-energy dependence of the potential depth. This is confirmed by the analysis
of \ref{ap:linearity}, which shows that, provided the potential is already much deeper than the binding energy, then the change in potential depth is of order the change in the binding energy. This suggests that although the NCSMC predictions have different features (see Table~\ref{TableNCSMC}), the \textit{ab initio} $\alpha$-$d$ binding potential, resulting from the many-body calculation,  is only slightly affected by these changes. 

Because the square well potential provides a simple explanation as to why the reduced partial widths stay constant in various \textit{ab initio} predictions, we now use a more realistic two-body potential to investigate the ANC-binding energy correlation. We consider a Woods-Saxon (WS) real potential
\begin{equation}
    V(r)=\frac{-V_R}{1+\exp [(r-R_R)/a_R]},
\end{equation} defined by its depth $V_R$, its radius $R_R=r_R\, 4^{1/3}$ and its diffuseness $a_R$. The diffuseness $a_R=0.65$~fm is fixed to a value commonly used in single-particle models. The depth of the potential is chosen to generate one radial node and reproduce various binding energies [-1.9,-1.3]~MeV.  The correlation between the binding energy and ANC then emerges from these calculations. 

In Fig.~\ref{fig:scaling_rmatrix}, we show this result as the dot-dashed line. It is shown for three radii, $r_R=1$~fm, $r_R=1.2$~fm and $r_R=1.4$~fm; this corresponds to a typical range of radii used in single-particle models~\footnote{We also note that the parameters $r_R=1.2$~fm and $a_R=0.65$~fm were found by Kubo and Hirata~\cite{Kubo1972} to accurately reproduce the $s$-wave $\alpha$-$d$ overlap function determined using a resonating-group method calculation of the ${}^6{\rm Li}$ ground state~\cite{Hasagawa1967}.}. The magnitude of these curves is adjusted to reproduce the ANC of the NCSMC prediction 9 (see Table~\ref{TableNCSMC}).
The correlation obtained in this simple model reproduces the one seen in the NCSMC calculations. For all $r_R$ considered,  the correlation line has less curvature than the one predicted by the phenomenological $R$-matrix.

\subsection{The interior norm, its linearity and a prediction for the ANC}

\label{subsec:perturbative}

To complement  our understanding of the correlation between the binding energy and  ANC$^2$ as well as to identify the relevant scales, we consider a perturbative approach. \ref{ap:linearity} shows that, when a small change is made to a  potential, then the corresponding change in the interior wave function is first order in the binding energy. The phenomena described thus far are therefore clearly generic to bound states that are ``fine tuned'', i.e., correspond to energies much smaller than the depth of the potential that formed them. 

The ``interior norm'' of the two-body state at a certain separation radius $\tilde{R}$, $I_{\rm in}(\tilde{R})$, was defined in Eq.~\eqref{eq:interiornorm}.
This is a scale dependent quantity, which really can only be interpreted as an interior norm if the value of $\tilde{R}$ is chosen close to the range of the potential.
That scale dependence 
for sufficiently large $\tilde{R}$ is entirely analytic, as long as the binding energy and ANC of the state are known. 
From the fundamental theorem of the calculus, 
\begin{equation}
    \frac{\partial I_{\rm in}}{\partial \tilde{R}}=|u_0(\tilde{R})|^2.
\end{equation}
For $\tilde{R} \rightarrow \infty$, $I_{\rm in}(\tilde{R}) \rightarrow N_0$, 
it follows that
\begin{equation}
    N_0=I_{\rm in}(\tilde{R}) + C_0^2 I_{\rm as}(\tilde{R}).
\end{equation}
 
\begin{figure}
    \centering
    \includegraphics[width=0.7\linewidth]{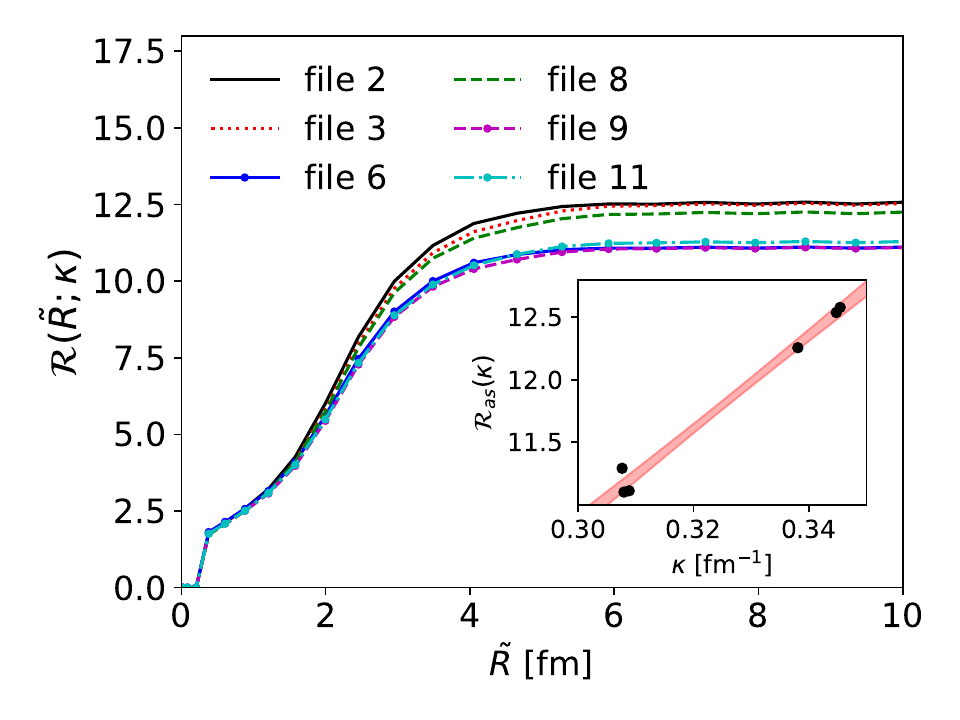}

    \caption{Ratio $\mathcal{R}(\tilde{R};\kappa)$~\eqref{eq:calR} obtained with various NCSMC calculations. The inset shows the asymptotic values  $\mathcal{R}_{as}(\kappa)$ as a function of the binding momentum $\kappa$. The red band corresponds to the $1\sigma $ uncertainty obtained from the linear regression of the NCSMC calculations for $\mathcal{R}_{as}(\kappa)$, evaluated at 10~fm.}
    \label{fig:fig1}
\end{figure}

 Let us define the ratio of two different calculations of the probability outside the radius $\tilde{R}$
 \begin{equation}
 {\cal R}(\tilde{R};\kappa) \equiv  \frac{N_0 - I_{\rm in}(\tilde{R})}{2 \kappa I_{\rm as}(\tilde{R})}.
 \label{eq:calR}
 \end{equation}
A plot of ${\cal R}(\tilde{R};\kappa)$ as a function of $\tilde{R}$ will then become independent of $\tilde{R}$ once $\tilde{R}$ is sufficiently large that the asymptotic region has been reached. 
Establishing where this happens is a novel way to assess the onset of the asymptotic regime for two-body bound states; it is especially pertinent for understanding where such behavior sets in for integrated quantities. Figure~\ref{fig:fig1}
 shows the ratio $\mathcal{R}(\tilde{R};\kappa)$ computed from  various NCSMC calculations. The asymptotic value is attained at $\tilde{R}=4$ fm for all calculations. If we denote the resulting asymptotic value as
\begin{equation}
    \lim_{\tilde{R} \rightarrow \infty} {\cal R}(\tilde{R};\kappa) \equiv {\cal R}_{\rm as}(\kappa),
\end{equation}
then
\begin{equation}
   \frac{C_0^2}{2 \kappa}={\cal R}_{\rm as}(\kappa).\label{eqAsKappa}
\end{equation}
 Extracting the values of ${\cal R}_{\rm as}(\kappa)$ for the different NCSMC calculations and plotting them as a function of $\kappa$ produces the result in the 
inset of Fig.~\ref{fig:fig1}. We see that---over this range of $\kappa$---${\cal R}_{\rm as}(\kappa)$ is approximately linear in $\kappa$.

This linearity can be derived by noting that the interior norm varies linearly with the binding energy, provided it is 
evaluated for $\tilde{R} \approx R_{pot}$, with $R_{pot}$ the range of the underlying potential {\it and} as long as the change in binding energy is small compared to the depth of the potential which is causing the binding. That is,
\begin{equation}
    |\delta I_{\rm in}(R_{pot})| \sim \frac{\hbar^2\kappa_0 \delta \kappa}{\mu V_0},
    \label{eq:linearity}
\end{equation}
with $\kappa_0$ some reference value of the binding momentum that is generated by a potential of depth $V_0$ and $\kappa=\kappa_0+\delta \kappa$. This follows directly from the linearity of changes in the interior wave function, see \ref{ap:linearity} for details.

Meanwhile, the denominator in Eq.~\eqref{eq:calR} can be expanded as a Taylor series around $\kappa=\kappa_0$. Since the integral can be expressed in terms of special functions, the derivative can be written out analytically, but the result is not particularly enlightening. The key point is that the convergence properties of this series are set by the expansion parameter $\kappa R_{pot}$\footnote{In principle there is a radius of convergence associated with the expansion of the Sommerfeld parameter around the value $k_C/\kappa_0$, but the singularities associated with this part of the expansion are much further away in this problem, since $\eta$ is markedly less than $1$.}, with $R_{pot}$ the range of the potential. %

\begin{figure}
    \centering
    \includegraphics[width=0.7\linewidth]{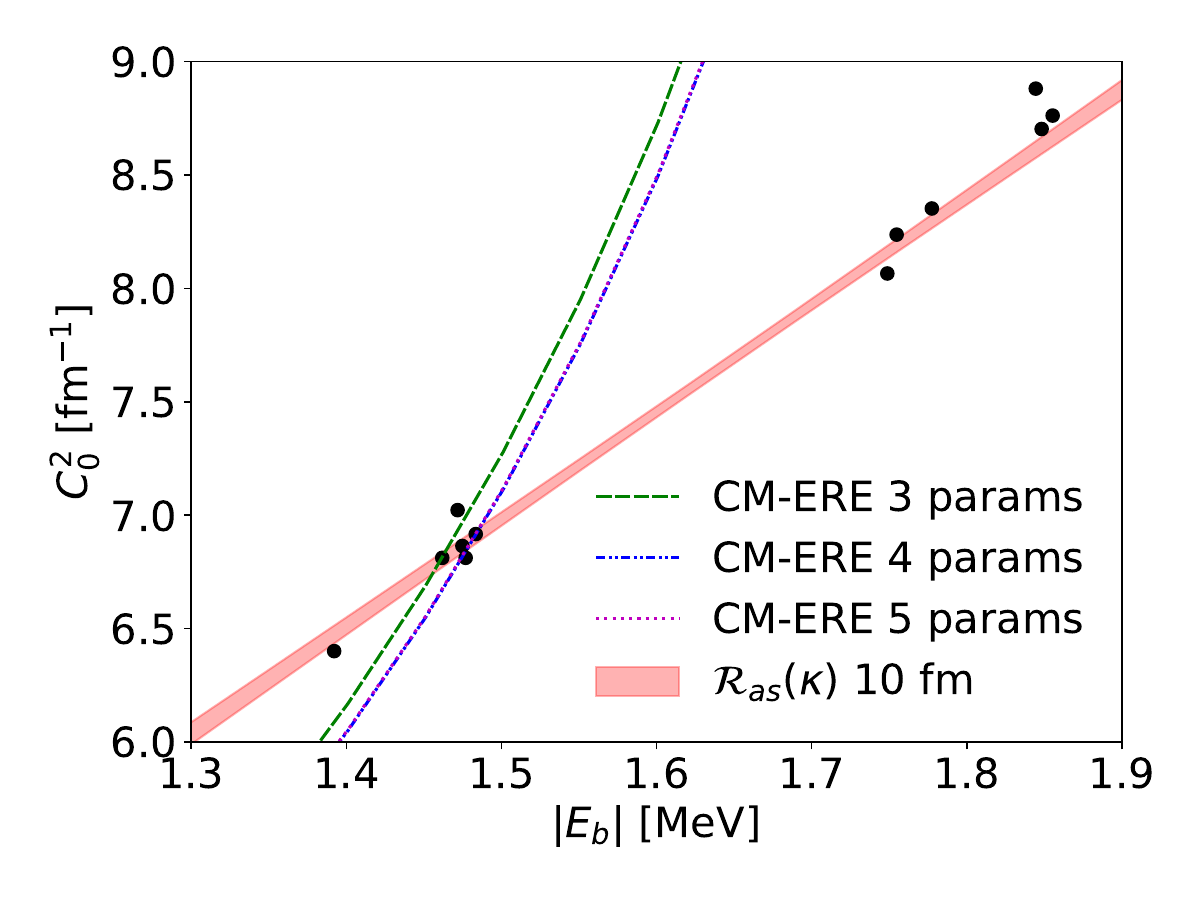}
    \caption{Comparison of the correlation between ANC$^2$s and the  binding energy obtained using  the perturbation expansion~\eqref{eq:prediction}  (red band) and CM-ERE. The  red band represents the $1\sigma$ uncertainty associated to the fit  of $R_{\rm as}(\kappa)$. }
    \label{fig:scaling_Rintegral}
\end{figure}

If we assume $\mu V_0 \sim \frac{\hbar^2}{R_{pot}^2}$
we have
\begin{equation}
    \frac{C_0^2}{2 \kappa} \approx {\cal R}_{\rm as} (\kappa_0) + (\kappa-\kappa_0) {\cal R}_{\rm as}'(\kappa_0) + {\cal O}[(\kappa-\kappa_0)^2],
    \label{eq:prediction}
\end{equation}
with ${\cal R}_{\rm as}'(\kappa_0) \sim R_{pot}\, {\cal R}_{\rm as}(\kappa_0)$.
Therefore the value of the ratio ${\cal R}(\kappa)$ in this scaling region should change linearly with $\kappa$ in some window around $\kappa_0$.  This linear behavior, i.e., the value of $\mathcal{R}_{\rm as}'(\kappa_0)$, can be determined by fitting to the asymptotic values  $\mathcal{R}_{\rm as}(\kappa)$ of various NCSMC calculations, that are shown in the inset of Fig.~\ref{fig:fig1}. 
Keeping only the linear dependence of~\eqref{eq:prediction}, i.e., truncating higher-order terms, leads to an accurate reproduction of the correlation between the ANC$^2$s  and the binding energy (red band in Fig.~\ref{fig:scaling_Rintegral}). 

The computation with the Woods-Saxon potential predicted a similar linear dependence in this range of $\kappa$. Compared to that calculation the one presented in this subsection is less predictive, since the slope of $\mathcal{R}_{\rm as}(\kappa)$ is fit to NCSMC data here. But it also shows that there is indeed nothing special about the Woods-Saxon potential, and any potential that is deep compared to the binding energy of the state will give a similar prediction. 

The expansion in $\kappa R_{pot} $ converges quickly; such a rapid convergence could be observed for other loosely-bound systems. 
This result can also be obtained using formulations that relate the ANC to an integral over the interior of the many-body wave function~\cite{PhysRevC.37.2859,PhysRevC.71.064305,Nollett:2011qf}, assuming the many-body matrix element's dependence on $\kappa$ is weak, and expanding the integrand in powers of $\kappa$ for small $\kappa$. 

Before closing this section we observe that the correlation seen in Fig.~\ref{fig:scaling_Rintegral} is {\it not} the one predicted by Halo EFT at leading order---or, equivalently, by the CM-ERE with only a scattering length in the effective-range function. Leading order Halo EFT predicts the correlation~\cite{Ryberg:2015lea}
\begin{equation}C_0^2=\frac{2 \kappa \Gamma^2(1+ \eta_b)}{2 \eta_b^2 \Psi'(\eta_b) - 1 - 2 \eta_b},\label{eq:LOHEFT}
\end{equation}
where $\Psi$ is the digamma function.
For the values of $\kappa$ and $\eta_b$ pertaining to Fig.~\ref{fig:scaling_Rintegral} Eq.~\eqref{eq:LOHEFT} yields values of $C_0^2$ that are a factor of ten smaller than the {\it ab initio} prediction are found. This indicates that the correlation between the binding energy and ANC$^2$ is not a manifestation of universality in $^6$Li. As we will see in the next section, range effects, i.e., effects beyond leading order in Halo EFT/the CM-ERE, are crucial here. 

\newpage
\section{The ANC-phase-shift connection}
\label{sec:ANCPSconnection}

We  now investigate the connection between the $s$-wave ANC and phase shift. It is well known that the ANC squared is proportional to the residue of the scattering amplitude at the bound state pole~\cite{Locher1978,Hammer:2017tjm}.  To obtain the ANC, one can therefore compute the scattering  amplitude at positive energy, and then perform an analytic continuation to the negative energy at which the bound state resides.  There are various methods that provide a connection between bound and scattering states, and they are typically used to infer ANCs relevant for astrophysics from phase shifts, e.g. Refs.~\cite{Blokhintsev2018,PhysRevC.96.034601,Ando2024}. One method is the phenomenological $R$-matrix described above; we will employ that method 
in Subsection~\ref{subsec:Rmatrixresults}. 

First though, in Subsec.~\ref{subsec:CMEREresults} we use the Coulomb-modified Effective Range Expansion (CM-ERE) to perform the analytic continuation from $E > 0$ to the bound-state pole.  This approach goes back to the 1950 paper of \cite{PhysRev.77.647}, where the $s$-wave ANC of the deuteron was predicted from $n$-$p$ scattering data. For the more complicated case of charged-particle scattering, see the work of Ref.~\cite{PhysRevC.29.349}. 

The NCSMC results discussed above give us a unique opportunity to investigate the connection between $s$-wave ANCs and $s$-wave phase shifts.  We have several calculations that produce different scattering phase shifts and different ANCs. Does analytic continuation to the bound-state pole based on the CM-ERE work equally well for all of them? And if so, at what order must we treat the effective range function in order for the extrapolation to be successful?  

Before carrying out such an analysis of the CM-ERE for the \textit{ab initio} phase shifts and the corresponding ANCs, we first detail the key equations needed to extract the ANC from the scattering amplitude using the CM-ERE.

\subsection{Coulomb-modified Effective-Range Expansion: A review}
\label{subsec:CMEREreview}

When considering a problem in which Coulomb and nuclear interactions are involved, the scattering amplitude has a complicated analytic structure and  it is common to introduce a function with simpler analytical properties~\cite{Sparenberg,Blokhintsev2018,Hamilton,Haeringen}. The $s$-wave scattering amplitude is
\begin{eqnarray}
F_0(k^2)&=&\frac{e^{2i\delta_0}-1}{2ik} \frac{e^{\pi\eta} e^{2i\sigma_0}}{ \Gamma^2(1+i\eta)}\\
    &=&\frac{1}{k_C({\rm cot}\,  \delta_0-i)} \frac{e^{2\pi\eta}-1}{2\pi}, 
\end{eqnarray}
where $k$ denotes the wavenumber 
and $\sigma_0$ is the $s$-wave Coulomb phaseshift.

If there is a bound state at energy $E_b $ of momentum $\kappa$, the amplitude $F_l$ has a pole at $k^2=-\kappa^2$. The ANC of this bound state can be obtained from the residue of this amplitude
\begin{equation}
    C_0=\Gamma(1+\eta_b)\left[-\frac{dF_0^{-1}}{dk^2}\Bigg|_{k^2=-\kappa^2}\right]^{-1/2}.
\end{equation}
$F_0$ is not analytic around $E=0$, but it is related to the Coulomb effective range function $K_0(k^2)$ via
\begin{equation}
    K_0(k^2)=\frac{1}{F_0(k^2)}+2 k_C h(\eta^2), \label{eq:Kl}
\end{equation}
where
\begin{equation}
    h(\eta)
=\psi(i\eta)-\ln (i\eta) +\frac{1}{2i\eta}.\end{equation} 
$K_0(k^2)$ is an analytic function of $k^2$ around $k^2=0$, with its radius of convergence set by the nearest singularity of the $\alpha$-$d$ scattering amplitude other than the two-body scattering cut and the left-hand cut associated with the Coulomb potential. It can therefore be Taylor expanded~\cite{Sparenberg,Blokhintsev2018}
\begin{equation}
K_0(k^2)=-\frac{1}{a_0} + \frac{r_0}{2} k^2 + \frac{\mathcal{P}_0}{4} k^4 + \mathcal{Q}_0 k^6 + \mathcal{R}_0 k^8 + \mathcal{S}_0 k^{10} +{O}(k^{12}). 
\end{equation}
Thus, if we can constrain the coefficients in this expansion, and hence $F_0$, at positive energies, i.e., the scattering regime, we can then perform an analytic continuation to negative energies and deduce the value of the ANC.

In this section, we use this approach to investigate the connection between the scattering phase shifts and the bound state properties. Our study is methodologically similar to 
that of Refs.~\cite{Yarmukhamedov:2011kd,Sparenberg}, in that we increase the order of the polynomial used for $K_0(k^2)$ until we find stability in the ANC.

\subsection{Results from CM-ERE}
\label{subsec:CMEREresults}

We analyze the phase shift data for the $\ell=0$ channel 
from the NCSMC calculations listed in Table~\ref{TableNCSMC}. Unless otherwise stated we consider phase shifts between 0.1~MeV and 3 MeV, and use data spaced by 0.05 MeV. The numerical errors on these results are negligible. 
We convert the phase shifts to the function $K_0(k^2)$ and fit the resulting data to polynomials of degree $1$, $2$, $3$, $4$, and $5$ in $k^2$. The fit of degree $n$ is hereafter labeled the fit of $O(k^{n+2})$, since that is the size of the first term it omits in $K_0(k^2)$. We employ a least-squares metric, but weight it as if we had a fractional error in $K_0(k^2)$~\footnote{The size of the fractional error does not matter since it does not affect the position of the $\chi^2$ minimum.}. This procedure is not statistically based but allows us to give some importance to the low-energy phase shifts, since $K_0$ becomes small at low energies. The $\chi^2$ is already well below the nominal number of degrees of freedom for the fit up to $O(k^8)$, i.e., including terms up to $k^6$, and the errors  on the effective-range parameters from the fit are negligible. 

\begin{figure*}
    \centering
    \includegraphics[width=\linewidth]{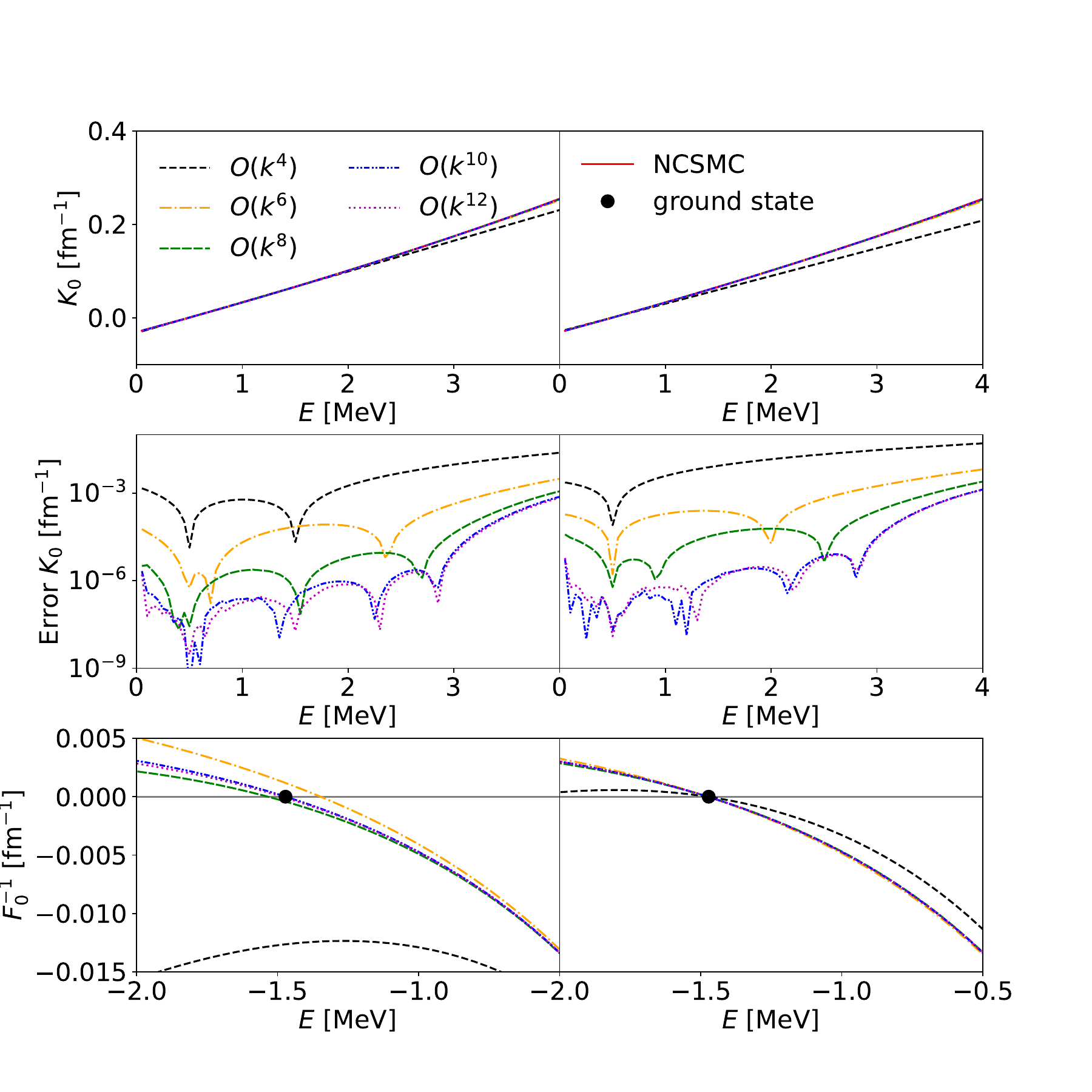}
    \caption{The $s$-wave phase shift  effective range function (top panel), the absolute error of the CM-ERE  (middle panel) and the inverse amplitude (bottom panel) from NCSMC calculation 9 and the CM-ERE for fits at various orders.  The point indicates the location of the bound state in the calculation. The left (right) plots correspond to the results without (with) the bound state pole position imposed.}
    \label{fig:phase_erange}
\end{figure*}

We note that the least-squares fit tends to produce large canceling coefficients at higher orders $O(k^{10})$ and $O(k^{12})$. This---together with the very small value of the $\chi^2$ found in the fit to $O(k^{12})$---makes the fit somewhat unstable. Care is required in order to ensure that the optimization algorithm converges to the minimum. 
The results of this procedure for NCSMC file 9 (bound state energy -1.48 MeV) are shown in the left-hand panel of Fig.~\ref{fig:phase_erange}. The top panel shows the fit itself, the second panel the residuals, and the third the extrapolation of $F_0^{-1}$ the bound-state pole. 
The fit $O(k^4)$ that includes only the effective range is poor, and the extrapolation does not even predict a bound state. The fit  $O(k^6)$ predicts a somewhat too-shallow bound state, and after that the convergence of the bound-state position with order is reasonable, although it continues to oscillate around the true value. There is no visible improvement in the residuals between the $O(k^{10})$ and $O(k^{12})$ fits. The binding energy obtained when the fit is done neglecting terms of  $O(k^8)$ is 1.54 MeV, 1.47 MeV for $O(k^{10})$, and 1.49 MeV for $O(k^{12})$. So, even though the fit at $O(k^{12})$ is not noticeably different from the fit at $O(k^{10})$, it still produces a shift of 0.02 MeV in the binding energy. 

Therefore we choose to  fix the position of the bound-state pole to the value predicted by the  NCSMC . This can be achieved  by imposing the constraint that $F_0^{-1}$ equals zero for $k^2=-\kappa^2$~\cite{Sparenberg}, which in turn can be achieved by expressing $1/a_0$ in terms of all the other coefficients and $\kappa$. We have therefore expanded $K_0(k^2)$ as a Taylor series around $-\kappa^2$, instead of around $0$. This constrained CM-ERE is more stable: the results are shown in the right panel of Fig.~\ref{fig:phase_erange}. It remains the case that the fit at $O(k^{12})$ is not an improvement over the $O(k^{10})$ one. By computing $\frac{dF_0^{-1}}{dk^2}$ at $k^2=-\kappa^2$ we then obtain the ANC$^2$. We find $6.51~{\rm fm}^{-1}$ at $O(k^6)$, $6.96~{\rm fm}^{-1}$ at $O(k^8)$, $6.81~{\rm fm}^{-1}$ at $O(k^{10})$, and $6.81~{\rm fm}^{-1}$ at $O(k^{12})$. The value at  $O(k^{12})$ is within 1\% of the true value of $6.86~{\rm fm}^{-1}$ found in the NCSMC ${}^6$Li ground-state wave function. 
 
Now we consider the case of File 7,  which leads to a separation energy of 1.84~MeV. Here we consider the CM-ERE extrapolation in which the bound-state pole is fixed. The extrapolation  produces a $C_0^2$ of 9.10, 8.58, and 8.65 (all in ${\rm fm}^{-1}$) at $O(k^8)$, $O(k^{10})$, and $O(k^{12})$, respectively. Given the large shift from $O(k^8)$ to $O(k^{10})$ and the fact that the $O(k^{12})$ fit is only marginally better it is unclear that the fit has stabilized. And indeed, the true ANC$^2$ is a few percent larger than the $ 8.65~{\rm fm}^{-1}$ number found via extrapolation: it is $8.88~{\rm fm}^{-1}$. The extrapolation from the scattering region does not work as well here, presumably because it is to a binding energy that is approximately 25\% larger than in the case of extrapolation to the physical binding energy.

Therefore extrapolation from the scattering regime to the bound-state regime can be successfully carried out using the CM-ERE, as long as the form that constrains the CM-ERE using the energy of the bound-state pole is employed, and the binding energy  is not too large. The CM-ERE extrapolant must include several terms if it is to be stable and accurate, but the fact the extrapolation works demonstrates that the {\it ab initio} NCSMC scattering amplitude has indeed the correct analyticity properties around the $\alpha$-deuteron threshold.

We now take the CM-ERE, written as an expansion of $K_0(k^2)$ around $-\kappa^2$, and vary $\kappa$ while keeping the coefficients in that expansion at the values obtained from the fit to the phase shifts of NCSMC calculation 9. This produces the green dashed curve for $O(k^8)$, the blue dash-dotted one for $O(k^{10})$ and purple dot-dashed  curve for $O(k^{12})$ in Fig.~\ref{fig:scaling_Rintegral}. The CM-ERE fails to predict the dependence of the ANC$^2$ on $\kappa$.  The CM-ERE can, if carried out to high enough order, predict the energy dependence of the amplitude for a given $\chi$EFT potential, but it cannot predict the correlation between $C_0^2$ and $\kappa$. That is because all the coefficients needed for the extrapolation---not just the $1/a_0$ term---are implicitly dependent on the details of the potential used to bind the system, and hence on the separation energy of ${}^6$Li. 

We also perform a  systematic analysis on all NCSMC calculations  in Fig.~\ref{fig:ere_rm_comparison}, which shows the ratio between the ANC$^2$ obtained with the CM-ERE analysis and the true value of the ANC$^2$. The green, blue and magenta circles correspond to the CM-ERE analysis for $O(k^8)$, $O(k^{10})$ and $O(k^{12})$, respectively. This figure will be discussed in more detail in Subsec.~\ref{subsec:RmatrixvsCMERE}. 

We will also briefly mention two other methods for fitting the CM-ERE that were investigated. The first is not actually fitting, but rather using a Lagrange interpolating polynomial to describe the CM-ERE, fixing it with the known value at the bound state and selected NCSMC phase shift points uniformly spaced over the chosen energy interval. This approach gave ANC results and convergence properties that were essentially equivalent to those obtained with fitting that are described above. The other method is a Pad{\' e} function, as has been used in previous investigations~\cite{PhysRevC.48.2390,PhysRevC.43.822}. The specific form used is a variable-order polynomial divided by a first-order polynomial and the methodology was otherwise as described above. We found that fits with 2 or 3 free parameters, i.e., with a lower-order polynomial in the numerator, gave more accurate results for the ANC than the polynomial fits with the same number of free parameters. However, when the number of free parameters was increased to 4 or 5, the Pad{\' e} results were essentially equivalent to the polynomial fits. These findings indicate that the Pad{\' e} function may have some advantages in the case of fitting experimental data, where the uncertainties are accounted for.

\newpage
\subsection{Results from $R$-matrix}
\label{subsec:Rmatrixresults}

The $R$-matrix fits use the same input phase shifts and assumed uncertainties in $K_0$ as did the CM-ERE. In order to implement the fractional error weighting, it is necessary to convert the calculated phase shifts to $K_0$~\eqref{eq:Kl}. The phenomenological $R$-matrix formalism of Ref.~\cite{PhysRevC.66.044611} was utilized. For the single channel case, the phase shift is given by
\begin{equation}
e^{2i\delta_0} = e^{-2i\phi_0} \left( 1+2i P_0
  \sum_{\lambda\nu} \gamma_\lambda A_{\lambda\nu} \gamma_\nu \right),
\end{equation}
where the sums over $\lambda$ and $\nu$ are from one to the number of levels and $A_{\lambda\nu}$ are elements of the level matrix $\bm{A}$ that is defined by its inverse
\begin{equation}
\begin{split}
(\bm{A}^{-1})_{\lambda\nu} &= (E_\lambda-E)\delta_{\lambda\nu}-
  \gamma_\lambda \gamma_\nu (S_0+iP_0) \\
  &+ \left\{ \begin{array}{ll}
  \gamma_\lambda^2 S_{0 \lambda} & \lambda=\nu \\
  \gamma_{\lambda} \gamma_\nu
  \frac{S_{0 \lambda}(E-E_\nu) - S_{0\nu}(E-E_\lambda)}
  {E_\lambda-E_\nu} & \lambda\neq \nu \end{array} \right. .
\end{split}
\end{equation}
The phase shift is defined by the $R$-matrix level energies $E_\lambda$, reduced width amplitudes $\gamma_\lambda$, and Coulomb functions evaluated for the energy $E$ and channel radius $a$
\begin{align}
P_0 &= \frac{ka}{F_0^2+G_0^2} , \\
S_0 &= ka\frac{F_0' F_0 + G_0' G_0}{F_0^2+G_0^2}, \mbox{~and} \\
e^{i\phi_0} &= \frac{G_0+iF_0}{\sqrt{F_0^2+G_0^2}} ,
\end{align}
where $F_0$ and $G_0$ are the regular and irregular Coulomb functions, respectively, and the $'$ indicates differentiation with respect to $kr$.
The notation $S_{0\lambda}$ is used to denote the $S_0$ evaluated for the level energy $E_\lambda$. For negative energies, $S_0$ is given by Eq.~\eqref{eq:shift}.

The $R$-matrix fits were performed with two to four energy levels. The energy of the lowest level was always fixed to the ${}^6{\rm Li}$ ground state energy. The remaining energy level(s) were used to describe the ``background'', i.e., the remaining slowly-varying part of the physical amplitude. These background level(s) represent the contributions of high-energy basis states. They are typically necessary in a phenomenological $R$-matrix analysis. The background levels also depend significantly on the choice of channel radius. The use of background levels in phenomenological $R$-matrix analyses is discussed further in Ref.~\cite[IV.F]{RevModPhys.89.035007}. 

The $R$-matrix fits with 
\begin{itemize}
    \item two free parameters were performed by varying the reduced width of the bound state and the reduced width of a single background level fixed at 20~MeV;
    \item three free parameters were performed by varying the reduced width of the bound state and the energy and reduced width of a single background level;
    \item four free parameters were performed by varying the reduced width of the bound state, the energy and reduced width of the first background level, and the reduced width of a second background level fixed at 100~MeV;
    \item five free parameters were performed by varying the reduced width of the bound state, the energy and reduced width of the first background level, the reduced width of a second background level fixed at 70~MeV, and the reduced width of a third background level fixed at 200~MeV.
    \end{itemize}

\begin{figure}
\centering
\includegraphics[width=0.7\linewidth]{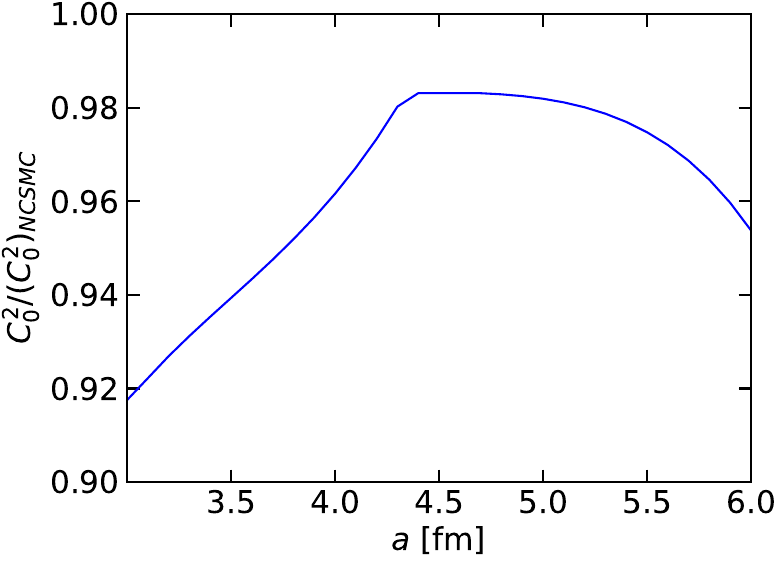}
\caption{The sensitivity of the ANC extracted from three-free-parameter $R$-matrix fits to the channel radius. The $y$-axis shows the ANC$^2$ compared to the NCSMC value. The phase-shift data used for the extrapolation to the bound-state pole were from NCSMC calculation 9.}
\label{fig:fit-rmatrix_a}
\end{figure}

The $R$-matrix fits are reasonably insensitive to the channel radius value used, provided $a$ is in the neighborhood of $4.5$ fm. The channel-radius sensitivity of the three-free-parameter $R$-matrix fits to NCSMC set~9 results using $E_{min}=0.1$~MeV and $E_{max}=3.0$~MeV is shown in Fig.~\ref{fig:fit-rmatrix_a}. The best fit is obtained for a radius of 4.4~fm, for which the extracted ANC$^2$ is within 2\% of its true value.  Relative to this minimum value obtained with $a=4.4$ fm, the $\chi^2$ is factor of 808 larger for $a=3.0$~fm and a factor of 12 larger for $a=6.0$~fm. We do not consider the channel radius to be a fit parameter; instead it should be chosen so that the extracted ANC does not change significantly if the radius is varied somewhat. In this case that fixes $a\in [4.4,5]$~fm. In this range, the ANC varies by less than 1\%, indicating that the ANC determination from phase shifts is  minimally sensitive to the channel radius. We thus adopt $a=4.5$ fm  in what follows. 

Figure~\ref{fig:ere_rm_comparison} shows the results of the $R$-matrix analyses of the phase-shift data from all NCSMC calculations. This figure is discussed in the next section.  
 \newpage

\subsection{Systematic analysis between CM-ERE and $R$-matrix approaches}\label{subsec:RmatrixvsCMERE}
\begin{figure}
\centering
\includegraphics[width=0.7\linewidth]{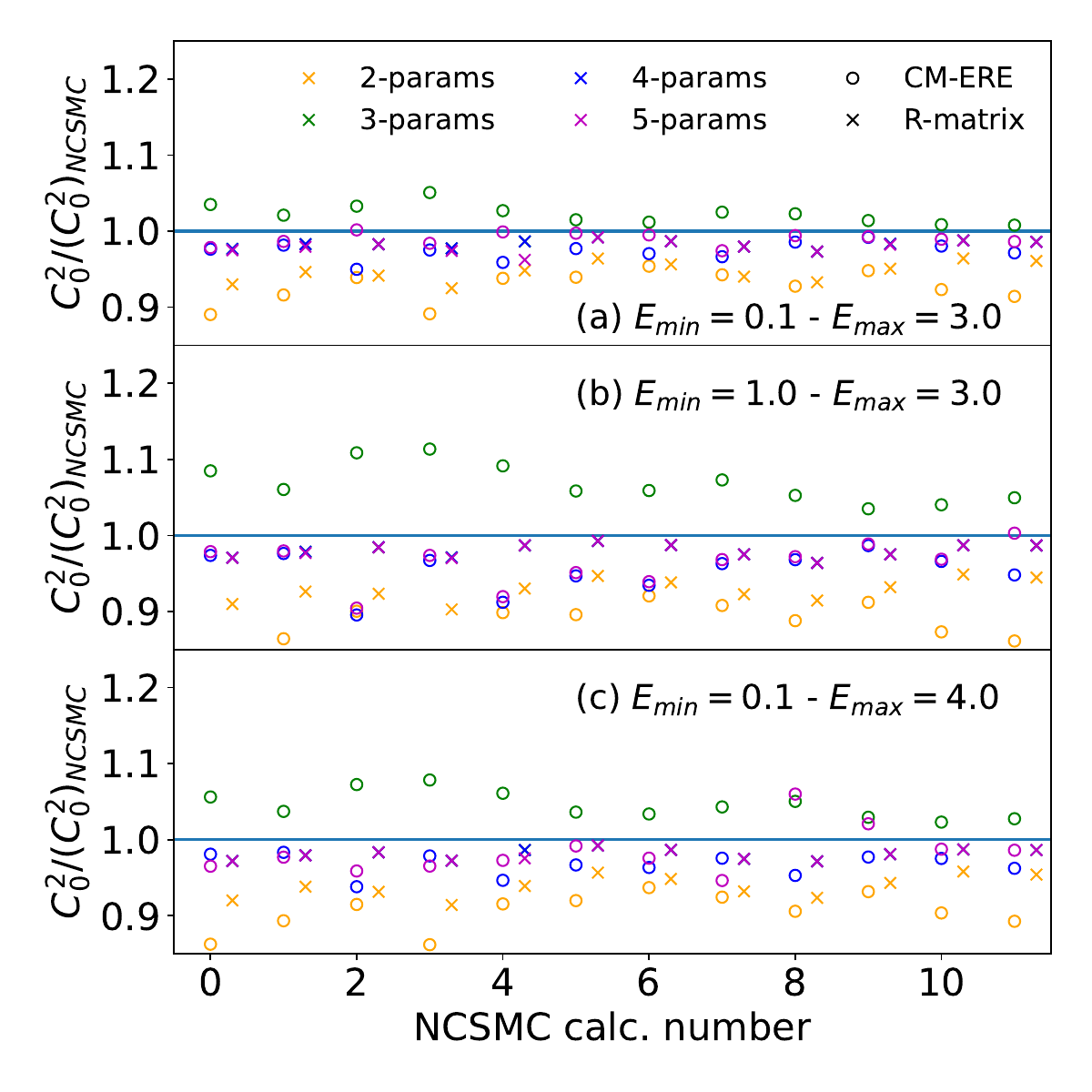}
\caption{Ratio of $C_0^2$ obtained with the CM-ERE (circles, Sec. \ref{subsec:CMEREresults}) and $R$-matrix (crosses, Sec. \ref{subsec:Rmatrixresults})  to the NCSMC value $C^2_{NCSMC}$ for each calculation in Table~\ref{TableNCSMC}. For each calculation we performed (constrained) CM-ERE and $R$-matrix fits with 2, 3, 4 and 5 parameters. Panel (a) corresponds to CM-ERE and $R$-matrix  fits of the Coulomb effective range function $K_0$ from  0.1 to 3 MeV, panel (b) to fits of $K_0$ from 1 to 3 MeV and panel (c)  to fits of $K_0$ from 0.1 to 4 MeV.}
\label{fig:ere_rm_comparison}
\end{figure}

Figure~\ref{fig:ere_rm_comparison} compiles the results for both CM-ERE (circles) and $R$-matrix (crosses) extractions of the ANC$^2$ for all NCSMC files (the $x$-axis corresponds to the calculation number). Results obtained with 2, 3, 4 and 5 parameters (\!$O(k^6)$, $O(k^8)$, $O(k^{10})$ and $O(k^{12})$ for the CM-ERE) are shown in, respectively, yellow, green, blue and magenta. The three panels show extrapolations that used three different phase-shift data sets: between 0.1 and 3~MeV (top panel), 1.0 and 3~MeV (middle panel) and 0.1 and 4 MeV (bottom panel). These results allow us to: 
\begin{enumerate}
    \item compare the results from the CM-ERE and $R$-matrix approaches,  
   \item analyze how large the discrepancy of each is with the true value, and if this discrepancy shows any systematic trends, 
   \item determine if one can naively quantify the uncertainties of the obtained ANC$^2$ by comparing the results of fits with different numbers of parameters,  and 
   \item study the impact of higher- and lower-energy phase shift data on the quality of the extrapolation.
   \end{enumerate}

First,  although both approaches are similar analytically~\cite{PhysRev.83.141,Humblet71},  the results obtained with both the $R$-matrix and CM-ERE obtained with various number of fitted parameters do not agree in almost all the cases studied here. By looking at all the different NCSMC calculations in all three panels, we noted two general features:  the discrepancy between both methods decreases when a larger number of parameters is included in the fit and increases for cases where the $^6$Li binding energy is larger than its physical value (file numbers 0, 2, 3, 4 and 7). 

\begin{figure}
\centering
\includegraphics[width=0.7\linewidth]{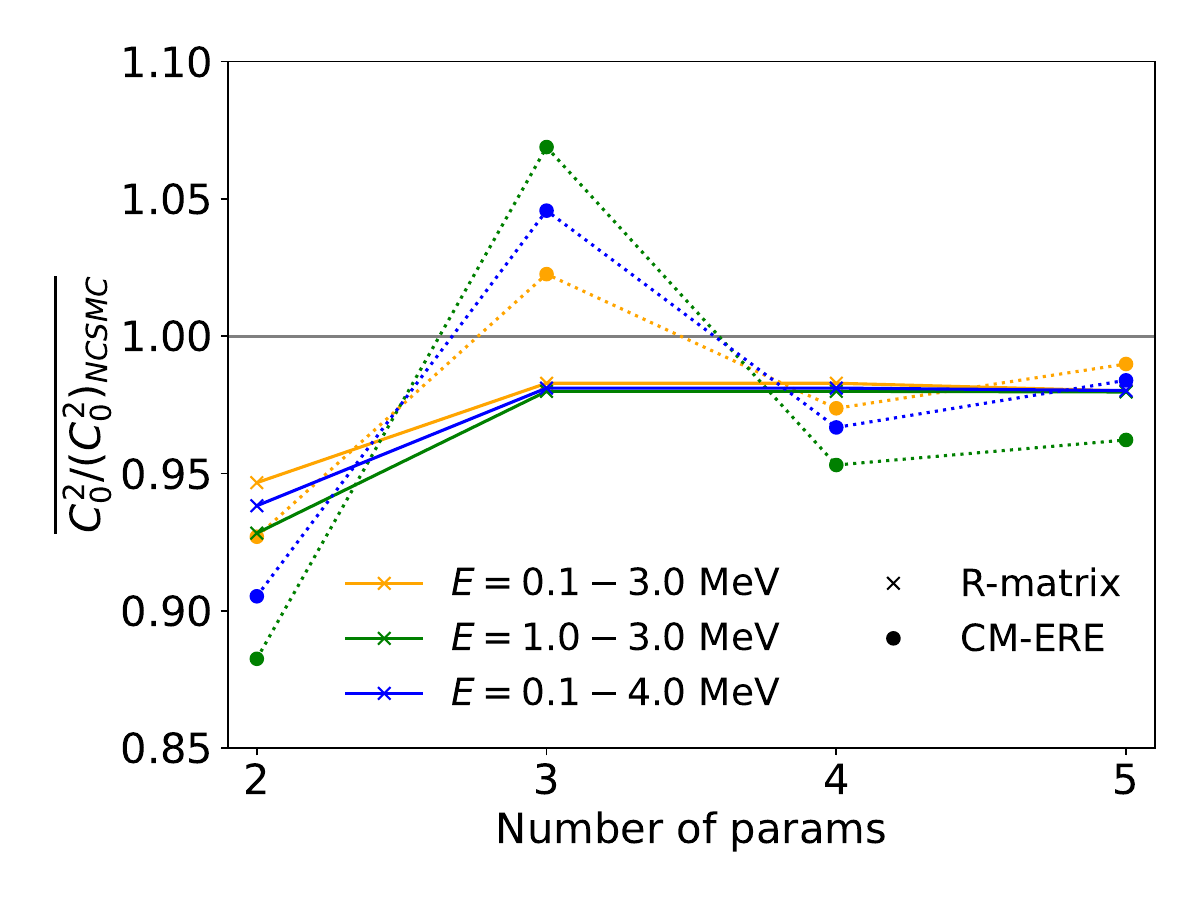}
\caption{ Ratio  of $C_0^2$ obtained with the CM-ERE (circles, Sec. \ref{subsec:CMEREresults}) and $R$-matrix (crosses, Sec. \ref{subsec:Rmatrixresults}) approaches  with its true value $C^2_{NCSMC}$ averaged over all NCSMC calculations (see Table~\ref{TableNCSMC}) for different numbers of parameters, as indicated in the horizontal axis. The yellow, green and blue lines correspond respectively to fits of $K_0$ from  0.1 to 3 MeV, from 1 to 3 MeV and  from 0.1 to 4 MeV.}
\label{fig:CompAvgRmatrixAndCMERE}
\end{figure}

Interestingly, the rate of convergence of the two approaches with the number of parameters is quite different. The ANC$^2$s extracted using the $R$-matrix converges faster than do the ones extracted using CM-ERE. 
This is  illustrated in Fig.~\ref{fig:CompAvgRmatrixAndCMERE}, which shows the ratio $C_0^2/(C_0^2)_{NCSMC}$, averaged over different NCSMC calculations for different number of fit parameters for both $R$-matrix (crosses, solid lines) and CM-ERE (circles, dotted lines).  For the $R$-matrix, there is little change to the ANC and very little improvement to the quality of the fit as number of free parameters is increased beyond three. This results because the multiple background levels at very high energies have very little effect on the energy dependence of the phase shift below 4~MeV. The reduced width parameters of these background levels are also highly correlated with each other. 
Compared to the $R$-matrix analysis, one needs to include a larger number of parameters in a CM-ERE analysis, although the error in the CM-ERE approach decreases in a regular way. 

Second, we note that, for all three panels in Fig.~\ref{fig:ere_rm_comparison}, an $R$-matrix fit with 3--5 parameters, or a CM-ERE fit with 5 parameters, approaches the true value. However, even fits that have apparently stabilized with order underestimate the true ANC$^2$ value by about 2\%. 
For both methods,  the discrepancy with the true value is larger for cases where the $^6$Li binding energy is larger than its physical value. This suggests that the determination of ANC from phase shift data  is more precise for loosely-bound states than deeply-bound ones. This observation is somewhat intuitive as a deeper bound state means that the gap  in energy between the phase-shift data and the bound state increases, requiring further extrapolation.

Third, to accurately determine  ANC$^2$s from phase-shift data, we must be able to assign a realistic error to the obtained value. In this work, we used  phase-shift data of $10^{-6}$ precision, hence no ``experimental" uncertainty needs to be taken into account.  One must quantify the uncertainties due to the truncation present in both theories, i.e. the truncation of the Taylor expansion of $K_0$ in the CM-ERE and the truncation in terms of poles included in the $R$-matrix expansion.  The simplest uncertainty quantification (UQ) method is to evaluate the uncertainty on $(C_0^2)^{(2n)}$ at $O(k^{2n})$ as  $(\Delta C_0^2)^{(2n)}=(C_0^2)^{(2n)}-(C_0^2)^{(2n-2)}$. However, for most cases this does not yield uncertainties with good coverage properties: the true value of the ANC$^2$ is not included in these error bars at the highest order employed. There are also systematic uncertainties from the neglect of the $d$-wave $\alpha$-$d$ channel and, in the case of $R$-matrix, the choice of channel radius. These considerations point out that a more sophisticated approach to uncertainty quantification needs to be devised in order for ANCs with reliable error bars to be determined from phase-shift data. We reserve this study for a future work.

Finally, we analyze the impact of low-energy  (resp. higher-energy) data by comparing in Fig.~\ref{fig:ere_rm_comparison} the top and the middle (resp. bottom) panels together and by comparing the yellow lines with the green (resp. blue) lines in  Fig.~\ref{fig:CompAvgRmatrixAndCMERE}.
For the CM-ERE, for all files, the smallest deviations from the true value are achieved for the fit window $E=0.1$--$3.0$ MeV. Removing low-energy data or adding higher-energy data increases the discrepancy between the obtained ANC$^2$ and the true value. 
This suggests that even in the ideal case considered here, i.e., no error on the phase shift and a fine energy mesh, low-energy phase shifts are important to precisely determine the ANC. 
The inclusion of the higher-energy data makes the  fit  less stable, as the cancellations between the high-order pieces of the expansion become more dramatic, driving the CM-ERE coefficients away from their true values.   
For the $R$-matrix, the situation is drastically different: as long as more than two parameters are used the accuracy of the $R$-matrix ANC extraction is only slightly diminished by increasing $E_{min}$ from 0.1 to 1.0~MeV, and increasing $E_{max}$ from 3.0 to 4.0~MeV has little impact.
 This impressive robustness suggests that one can use the $R$-matrix to extract ANCs reliably from relatively high-energy scattering data. The minimum energy that needs to be included in the data will likely vary in each case, and might be strongly dependent on the accuracy of the scattering data.

\section{Summary and outlook}
\label{sec:summary}

ANCs  play a key role in low-energy reactions, including radiative capture reactions relevant for astrophysics. In this work, we study how ANCs correlate with other bound-state observables and investigate if they can be reliably inferred from scattering data. To answer these two questions, this study analyzes the \textit{ab initio} NCSMC calculations of $^6$Li carried out in Ref.~\cite{PhysRevLett.129.042503} using various $\chi$EFT Hamiltonians and model spaces.  

 The first part of this work focuses on understanding the correlation between the $s$-wave  ANC$^2$ of  the $^6$Li ground state and its binding energy. We investigate this correlation using three different theoretical approaches: $R$-matrix, a single-particle model, and a perturbative expansion. We are able to explain this correlation assuming that the changes in the  $\alpha$-$d$ relative potentials between the various NCSMC calculations, and hence the changes in the $^6$Li binding energy, are small compared to the depth of the central potential. This suggests that the different NCSMC calculations lead to similar $\alpha$-$d$ potentials. This is a non-trivial observation since these are  results from complex six-body calculations using different Hamiltonians and model spaces.

 If such correlations are present for systems other than $^6$Li, they could be a useful tool for the evaluation of \textit{ab initio }capture rates relevant for astrophysics. Although this is an interesting avenue of investigation, several points  need to be   clarified. In this study, we used a single-channel description of the $\alpha$-$d$ system. Verification that such analysis can be generalized to coupled-channel problems is needed. A systematic analysis of various systems to determine when such correlations are expected and observed would also be useful.

 In the second part of this work, we explored how accurately one can extract ANCs from phase shifts. We use the \textit{ab initio} phase shift prediction as an ideal data set: they include low-energy data (down to 0.01 MeV),  have a fine energy mesh, and negligible uncertainties. 
 Using this ideal dataset, we verify that one can extract ANCs using two different methods: CM-ERE and $R$-matrix.  
Although the $R$-matrix result can be analytically mapped to the CM-ERE~\cite{PhysRev.83.141,Humblet71}, the two approaches have a different convergence rate with the number of parameters included in the fit and only agree for a fairly large number of parameters. On the one side, $R$-matrix converges quickly: with only a three-parameter fit. It also shows a limited sensitivity to the channel radius. On the other side, the CM-ERE requires as many as five parameters to converge and the fitting procedure can be  unstable.

Furthermore, both CM-ERE and R-matrix methods respond differently to the inclusion  of higher-energy data ($>3$ MeV)  and the absence  of low-energy data ($<1$ MeV) in the calibration of their parameters. The  $R$-matrix approach is  robust to these changes in the dataset, providing consistent ANC values when low-energy data is removed or high-energy data as added. However,  the ANCs extracted with the CM-ERE becomes less accurate when low-energy data are missing, or  higher-energy data are included. This instability of the CM-ERE is likely due to the fact that at high orders its expansion involves canceling terms, which can take very larger values for high energy. 
 These tests  suggest that ANCs extracted with CM-ERE might be biased by the energy range of the dataset: ANC extractions using $R$-matrix are likely more accurate.

The quantification of uncertainties for the extracted ANCs is a subject for future investigation. Naive uncertainty quantification based on comparison of extracted ANCs using $R$-matrix and CM-ERE approaches and various number of parameters does not yield error bars with good coverage properties. This is perhaps due to unquantified model uncertainties stemming from the simple  single-channel description of the $\alpha$-$d$ system and the neglect of the $\alpha$-$n$-$p$  channel that opens at 2.2~MeV above the $\alpha$-$d$ threshold.
This motivates the development of more elaborate UQ schemes, a topic which we  reserve  for future work, where we will consider a simulated data set including realistic experimental errors. 

\section*{Acknowledgments}
This project was initiated at the KITP program ``Living near Unitarity'' and was therefore supported in part by grant NSF PHY-2309135 to the Kavli Institute for Theoretical Physics (KITP). It
also received financial support from the CNRS through the AIQI-IN2P3 project (CH), from the U.S. Department of Energy under grants no. DE-NA0004247 (CRB), DE-FG02-88ER40387 (CRB), and DE-FG02-93ER40756 (DRP), and via a Tage Erlander Professorship funded by the Swedish Research Council (Grant No. 2022-00215). This work is based in part on prior results obtained via the LLNL Computational Grand Challenge program.

\appendix

\section{Derivation of linear variation of interior norm with changes in binding momentum}

\label{ap:linearity}

We consider here a  s-wave bound state of two charged particles, of binding energy $B_0$ and binding momentum $\kappa_0=\sqrt{2\mu |B_0|}{\hbar}$, generated by a potential $V_0 f(r)$ that is confined to the region $r<R_{pot}$. The radial bound-state wavefunction of this system is denoted $u_0(r)$. 
Suppose that we now increase $V_0$ by an amount $\delta V_0$, leading to a state bound by $B=B_0+\delta B$. To first order in perturbation theory, the change in the binding energy is
\begin{equation}
    \delta B\approx-\delta V_0 \int_0^{R_{pot}} dr \, |u_0(r)|^2 f(r).
\end{equation}
Let us also define the change in the binding momentum $\delta \kappa$ with $\kappa=\kappa_0+\delta \kappa$. Using $\kappa = \sqrt{2\mu B}/\hbar$, one can write at first order in perturbation theory
\begin{equation}
      \delta \kappa\approx-\frac{ \mu \,\delta V_0}{\hbar^2 \kappa_0} \int_0^{R_{pot}} dr \, |u_0(r)|^2 f(r).\label{eq:deltakdeltaV}
\end{equation}
We also define the new wavefunction $u(r)=u_0(r)+\delta u(r)$. The change in the wave function $\delta u(r)$ can be obtained at first order in perturbation theory as
\begin{equation}
    \delta u(r)\approx-\frac{2 \mu}{\hbar^2} \,\delta V \int_0^{R_{pot}} dr' \, G_{i}(r,r';-B_0) f(r') u_0(r'),
    \label{eq:firstorderdeltau}
\end{equation}
where $G_i(E)=[E-\hat T-\hat{V}_0]^{-1}$ is the Green's function of the unperturbed system, defined using the kinetic and potential energy operators $\hat{T}$ and $\hat{V}_0$. In Eq.~\eqref{eq:firstorderdeltau} this Green's function is evaluated at the binding energy $-B_0$ of the unperturbed system. The integral need only be carried out from $0$ to $R_{\rm pot}$ because that is where $f(r')$ is non-zero.

The change in  the interior norm~\eqref{eq:interiornorm}
 due to this change in the potential is
\begin{eqnarray}
\delta I_{\rm in}(\tilde{R})\approx - \frac{2 \mu}{\hbar^2} \delta V_0 \int_0^{\tilde{R}} dr \,u_0 (r) 
\int_0^{R_{pot}} dr'   G_{i}(r,r';-B_0) f(r') u_0(r').
\end{eqnarray}
Using \eqref{eq:deltakdeltaV}, this change in the interior norm can be written as
\begin{eqnarray}
    \delta I_{\rm in}(\tilde{R})=2 \kappa_0 \delta \kappa \,\frac{\int_0^{\tilde{R}} dr \, u_0 (r)\int_0^{R_{pot}} dr'  G_{i}(r,r';-B_0) f(r') u_0(r')}{\int_0^{R_{pot}} dr \, |u_0(r)|^2 f(r)}.
\end{eqnarray}
Now we exploit the fact that $G_i$ is an operator that only connects points $r$ and $r'$ that are separated by distances of order $\frac{\hbar}{\sqrt{2 \mu V_0}}$, which also sets the typical size of this one-dimensional Green's function~\footnote{This assumes that regions where $f(r) \ll 1$ contribute negligibly to the integral.}. The size of the integral in the numerator is then 
\begin{eqnarray}
&&\int_0^{\tilde{R}} dr\,u^\dagger_0(r) \int_0^{R_{pot}} dr' \,  G_{i}(r,r';-\kappa_0^2) f(r') u_0(r') \nonumber\\
&&\hspace{6cm}\sim \frac{\hbar^2}{2 \mu V_0} \int_0^{\tilde{R}} dr \, |u_0(r)|^2 f(r).
\end{eqnarray}

Therefore
\begin{equation}
    \delta I_{\rm in}(R_{pot}) \sim \frac{\hbar^2 \kappa_0 \delta \kappa}{\mu V_0},
\end{equation}
as claimed in Eq.~\eqref{eq:linearity} above.
The interior norm $I_{\rm in}(\tilde{R})$ will change linearly as the binding momentum $\kappa$ changes. Indeed, the perturbation to the unnormalizaed wave function is small as long as $\kappa_0$ is small compared to the effective momentum in the potential well. 

\section*{References}
\bibliography{biblio}
\end{document}